\documentclass[conference]{IEEEtran}
\usepackage{fancybox}

\usepackage{cite}

\ifCLASSINFOpdf
   \usepackage[pdftex]{graphicx}
\else
   \usepackage[dvips]{graphicx}
\fi
\usepackage[cmex10]{amsmath}
\usepackage{array}
\usepackage{fixltx2e}

\usepackage{stfloats}

\usepackage{url}

\begin{document}
%
\title{Autonomous Spacecraft Navigation \\ Based on Pulsar Timing Information}



%
\author{\IEEEauthorblockN{Mike Georg Bernhardt\IEEEauthorrefmark{1},
Werner Becker\IEEEauthorrefmark{1}\IEEEauthorrefmark{2}, 
Tobias Prinz\IEEEauthorrefmark{1},
Ferdinand Maximilian Breithuth\IEEEauthorrefmark{1} and 
Ulrich Walter\IEEEauthorrefmark{3} \\[1ex]}
\IEEEauthorblockA{\IEEEauthorrefmark{1}Max-Planck-Institut f\"ur extraterrestrische Physik,\\ Gie\ss{}enbachstr. 1, 85741 Garching, Germany \\ Email:  mbernhardt@mpe.mpg.de \\[1ex]}
\IEEEauthorblockA{\IEEEauthorrefmark{2}Max-Planck-Institut f\"ur Radioastronomie,\\ Auf dem H\"ugel 69, 53121 Bonn, Germany \\[1ex]}
\IEEEauthorblockA{\IEEEauthorrefmark{3}Institute of Astronautics, Technische Universit\"at M\"unchen, \\ Boltzmannstr. 15, 85748 Garching, Germany}
}


\maketitle

\begin{abstract}
We discuss the possibility of an autonomous navigation system for spacecraft that is based on pulsar timing data. Pulsars are rapidly rotating neutron stars that are observable as variable celestial sources of electromagnetic radiation. Their periodic signals have timing stabilities comparable to atomic clocks and provide characteristic temporal signatures that can be used as natural navigation beacons, quite similar to the use of GPS satellites for navigation on Earth. By comparing pulse arrival times measured on-board the spacecraft with predicted pulse arrivals at some reference location, the spacecraft position can be determined autonomously with accuracies on the order of 5~kilometres. For a spacecraft at a distance of 10 astronomical units from Earth (e.g., Earth--Saturn), this means an improvement by a factor of 8 compared to conventional methods. Therefore this new technology is an alternative to standard navigation based on radio tracking by ground stations, without the disadvantages of uncertainty  increasing with distance from Earth and the dependence on ground control.

\end{abstract}


%
\IEEEpeerreviewmaketitle

\thisfancyput(2.2cm,-24cm){\footnotesize Presented at the 2nd International Conference on Space Technology, 15--17 Sept. 2011, Athens, Greece.}

\section{Introduction}

The standard method of navigation for interplanetary spacecraft is a combined use of radio data, obtained by tracking stations on Earth, and optical data from an on-board camera during encounters with solar system bodies. Radio measurements taken by ground stations provide very accurate information on the distance and the radial velocity of the spacecraft with typical random errors of about 1~metre and 0.1~millimetres per second, respectively \cite{madde}. However, the components of position and velocity perpendicular to the Earth-spacecraft line are subject to much larger errors, which are due to the limited angular resolution of the radio antennas. Interferometric methods can improve the angular resolution to about 25~nanoradians, corresponding to uncertainty in the spacecraft position of about 4~kilometres per astronomical unit of distance between Earth and spacecraft \cite{james}. With increasing distance from Earth, the position error increases as well, e.g., reaching a level of uncertainty in the order of 120~kilometres at the orbit of Neptune. Nevertheless, this technique has been used successfully to send space probes to all planets in the solar system and to study asteroids and comets at close range. However, it might be necessary for future missions to overcome the disadvantages of this method, namely the dependence on ground-based control and maintenance and the high uncertainty at large distances. Therefore, it is desirable to automate the procedures of orbit determination and orbit control in order to support autonomous space missions.

Possible implementations of autonomous navigation were already discussed in the early days of space flight \cite{battin}. In principle, the orbit of a spacecraft can be determined by measuring angles between solar system bodies and astronomical objects, e.g., the angles between the Sun and two distant stars and a third angle between the Sun and a planet. However, because of the limited angular resolution of on-board star trackers and sun sensors, this method yields spacecraft positions with uncertainties that typically accumulate to several thousand kilometres \cite{battin}. Alternatively, the navigation fix can be established by observing multiple solar system bodies: it is possible to triangulate the spacecraft position from images of asteroids taken against a background field of distant stars. This method was realized and flight tested on NASA's \mbox{Deep-Space-1} mission between October 1998 and December 2001. The Autonomous Optical Navigation (AutoNav) system on-board Deep Space~1 provided the spacecraft orbit with $1 \sigma$-errors of 250~kilometres and 0.2~metres per second \cite{riedel}. Although AutoNav was operating within its validation requirements, the resulting errors were relatively large compared to ground-based navigation.

An alternative approach to autonomous spacecraft navigation is based on pulsar timing information \cite{downs, chester, esa, sheikh, sheikh2}. 
In the following, we will explain the concept of a pulsar and describe the principles of pulsar-based navigation, with a special focus on the use of X-ray pulsars. Furthermore, we will discuss the achievable accuracy and the impact of new technological developments on the feasibility of pulsar-based navigation.

%
%

\section{Pulsars as Navigation Beacons}

Pulsars were first discovered in the late 1960s as sources of radio emission pulsating with very regular periods. Although the mechanisms of radio emission are still poorly understood, the most plausible explanation for this phenomenon is the lighthouse model: a pulsar is a rapidly rotating neutron star whose strong magnetic field produces conical beams of electromagnetic radiation along its magnetic axis. If the magnetic axis is tilted with respect to the spin axis of the star, pulses of radiation are observed as the beams sweep across our line of sight once or twice per rotation cycle (Fig.~\ref{fig1}). To date, almost 2000 pulsars have been found not only in the radio, but also the optical, X-ray and gamma-ray regimes of the electromagnetic spectrum \cite{atnf, becker, beckertruemper}. Three classes of pulsars can be distinguished according to their source of electromagnetic energy:

\begin{figure}[!t]
\centering
\includegraphics[width=2.4in]{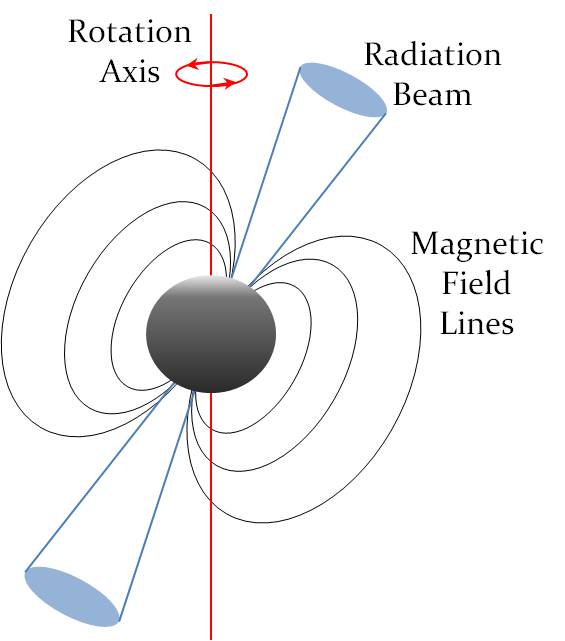}
\caption{Schematic illustration of a pulsar.} 
\label{fig1}
\end{figure}

1) \textit{Rotation-powered pulsars} radiate at the expense of rotational energy: The pulsar spins down as rotational energy is radiated away. There are two types of rotation-powered pulsars: \textit{Normal pulsars} have periods between tens of milliseconds to several seconds. They constitute more than 90~\% of the total pulsar population \cite{lyne}. About 7~\% of all pulsars fall into the category of \textit{millisecond pulsars}, which are usually defined to have periods in the range of about 1.5 to 30~milliseconds \cite{lyne} and exhibit very low spin-down rates. The majority of millisecond pulsars are found in binary systems. It is assumed that they are born with longer spin periods, but are spun up due to transfer of angular momentum from a companion star during a phase of accretion.

2) \textit{Accretion-powered pulsars} are close binary systems in which a neutron star is accreting matter from the companion star, thereby gaining energy and angular momentum. The observed pulses from an accretion-powered pulsar are due to the changing viewing angle of hot-spots on its surface. In contrast to rotation-powered pulsars, the spin behaviour of accretion-powered pulsars can be very complicated. They often show an unpredictable evolution of rotation period, with erratic changes between phases of spin-down and spin-up \cite{ghosh}.

3) \textit{Magnetars} are neutron stars with rather long rotation periods of a few seconds 
and exceptionally high magnetic fields of up to $10^{15}$ Gauss that are responsible for super-strong bursts of X-rays and gamma-rays. Their steady, ``quiescent'' emission component is powered either by the decay of the magnetic field \cite{duncan} or by accretion from a fall-back disk, which is a remnant of the supernova explosion \cite{truemper}.

Concerning their application to spacecraft navigation, the group of rotation-powered pulsars, especially the isolated millisecond pulsars, are particularly interesting for two reasons: First, they provide characteristic temporal signatures that can be used as natural navigation beacons; second, the stability of their rotation frequencies is comparable to, or even better than, the timing stability of atomic clocks \cite{matsakis}, which is most important, because the temporal resolution is the limiting factor to the performance of this navigation technique.

A pulsar-based navigation system can be designed for any energy band of the electromagnetic spectrum, but X-rays are preferable for several reasons \cite{ray, emadzadeh}: Whereas radio pulsars are usually very faint and, therefore, are observed by large radio antennas with diameters of typically 50 to 100~metres, X-ray telescopes can be built relatively compactly (see next chapter). Furthermore, radio waves are subject to interstellar dispersion, an effect that degrades the temporal resolution of pulsar data due to pulse smearing \cite{lorimer}. In contrast, X-ray 
propagation through the interstellar medium does not affect pulsar timing.

\begin{figure}[!t]
\centering
\includegraphics[width=3.5in]{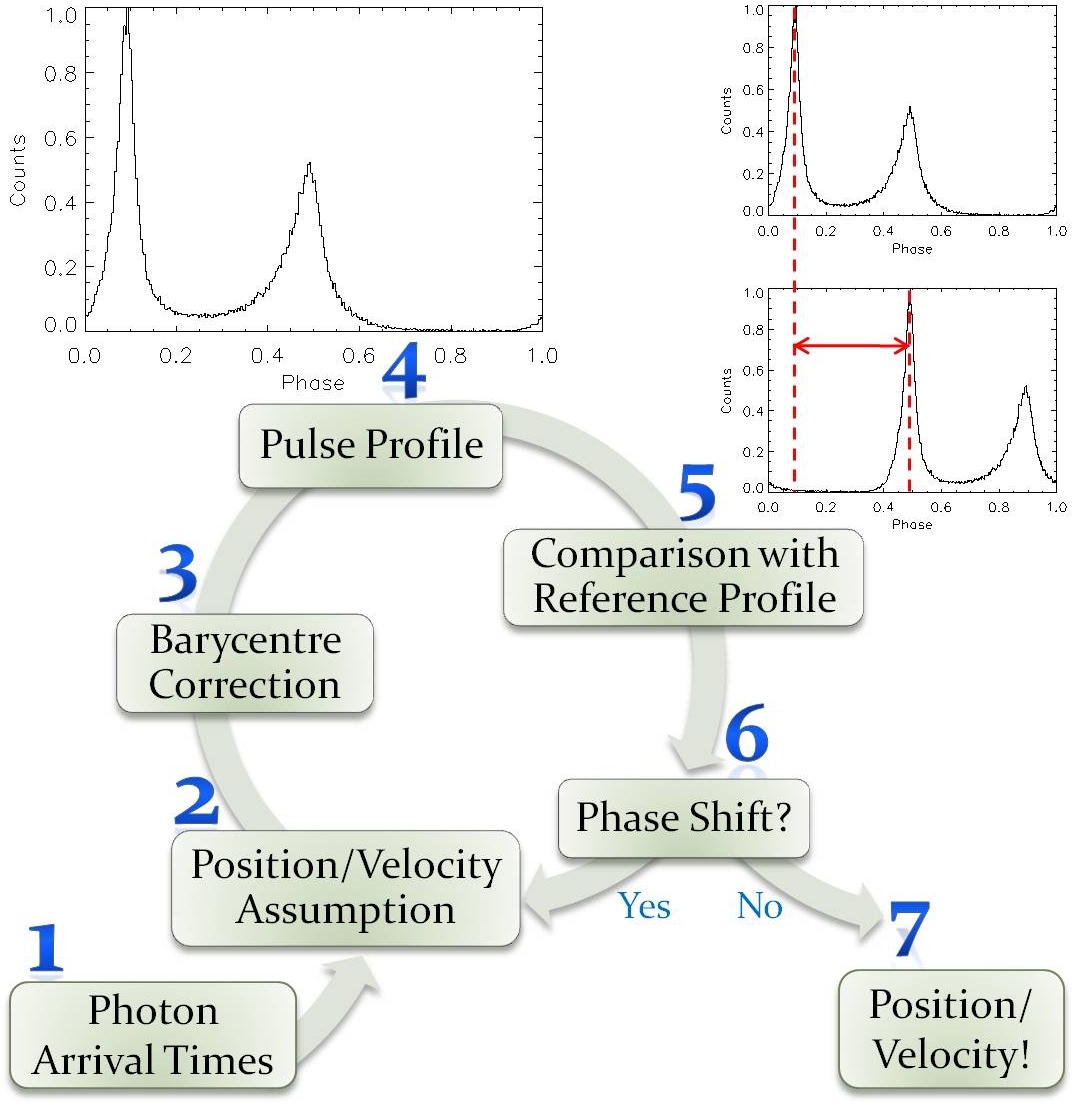}
\caption{Principle of pulsar-based navigation.}
\label{fig2}
\end{figure}

The principle of pulsar-based navigation is illustrated in Fig.~\ref{fig2}: Position and velocity of the spacecraft are determined in an iterative process, starting with the detection of individual X-ray photons emitted by a pulsar (1). The photon arrival times have to be corrected for the proper motion of the detector by transforming to an inertial reference location, e.g., the barycentre of the solar system. This correction requires a position and velocity assumption (2). The corrected arrival times (3) allow us to produce a pulse profile (4) that represents the temporal emission characteristics of the pulsar. Comparison with a reference profile (5) yields the pulse arrival at the barycentre (i.e., the phase of the global maximum in the pulse profile). A phase shift (6) with respect to the predicted pulse arrival at the barycentre corresponds to a range difference along the line of sight towards the observed pulsar. When the phase shift is zero, the assumption about position and velocity is correct (7). Eventually, a three dimensional position fix can be derived from observations of at least three different pulsars. Since the position of the spacecraft is deduced from the phase of a periodic signal, ambiguous solutions may occur. This problem can be solved by constraining the domain of possible solutions to a finite volume around an initial position assumption, or by observing additional pulsars.

In a previous study, we examined the achievable accuracy of a navigation system based on X-ray pulsars \cite{bernhardt}. We re-analysed all pulsar timing data from the X-ray satellites XMM-Newton, Chandra and the ROSSI X-ray Timing Explorer, and derived the temporal characteristics and typical errors of pulse arrival measurements for a set of $\approx$~60 X-ray pulsars \cite{prinz}. Using these data for spacecraft navigation, we found that position determination in three dimensions can be achieved with typical random errors of about 5 to 10~kilometres. The precision of a pulsar-based navigation system, though, strongly depends on the choice of pulsars and their spatial arrangement.

\section{X-Ray Telescopes for Navigation}

The design of an X-ray telescope suitable for navigation will be a compromise between angular resolution, collecting area and weight of the system. The currently operating X-ray observatories XMM-Newton and Chandra have large collecting areas (XMM 0.43~m$^2$, Chandra 0.08~m$^2$ effective area at 1~keV \cite{friedrich}) and attain very good angular resolution of less than 1 arcsecond (Chandra), but the focusing optics and support structures require a large mass, which is prohibitive for a navigation system. In recent years, ESA has put a lot of effort into the development of low-mass X-ray mirrors which can be used as basic technology for future large X-ray observatories and small planetary exploration missions. These light-weight mirrors are also very interesting for pulsar-based navigation.

\begin{figure}[!t]
\centering
\includegraphics[width=3.5in]{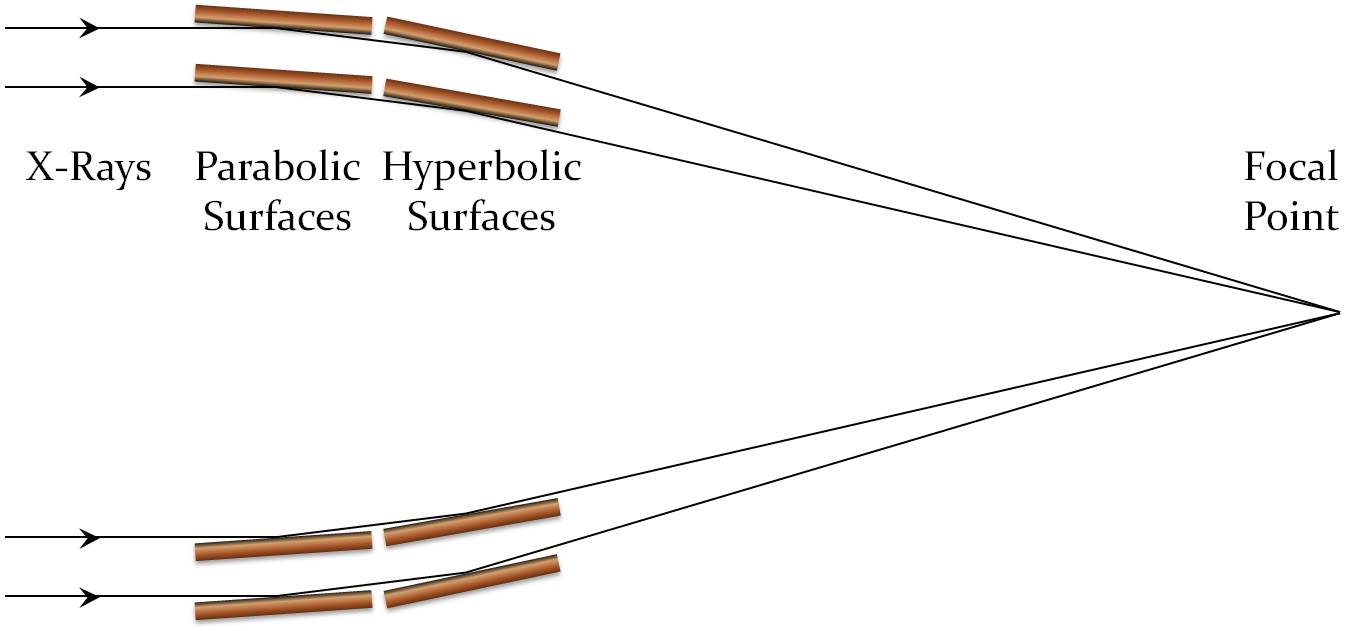}
\caption{A cross section through two nested pairs of mirrors illustrating the principle of grazing incidence reflection and focusing of X-rays.}
\label{fig3}
\end{figure}

A typical high-resolution X-ray telescope uses focusing optics based on the Wolter-I design \cite{wolter}. The incoming X-ray photons are reflected under small angles of incidence in order not to be absorbed and are focused by double reflection off a parabolic and then a hyperbolic surface (Fig.~\ref{fig3}). This geometry allows nesting several concentric mirror shells in order to increase the collecting area and thereby improve the sensitivity.

A novel approach to X-ray optics is the use of pore structures in a Wolter-I configuration \cite{bavdaz1, beijersbergen1, bavdaz2}. X-ray photons that enter a pore are focused by reflections on the walls inside the pore. 
In contrast to traditional X-ray optics with separate mirror shells that are mounted to a support structure, pore optics form a monolithic, self-supporting structure that is light-weight, but also very stiff and contains many reflecting surfaces in a compact assembly. Two different types of pore optics have been developed, based on silicon and glass (Table~\ref{table1}):


1) \textit{Silicon Pore Optics} \cite{collon1, ackermann, collon2} use commercially available and mass-produced silicon wafers from the semiconductor industry. These wafers have a surface roughness that is sufficiently low to meet the requirements of X-ray optics. A chemo-mechanical treatment of a wafer results in a very thin membrane with a highly polished surface on one side and thin ribs of very accurate height on the other side. Several of these ribbed plates are elastically bent to the geometry of a Wolter-I system, stacked together to form the pore structure and finally integrated into mirror modules. Silicon Pore Optics are intended to be used on large X-ray observatories that require a small mass per collecting area (in the order 200 kg/m$^2$) and angular resolution of about 5 arcseconds or better.

2) \textit{Glass Micropore Optics} \cite{beijersbergen2, collon3, wallace} are made from polished glass blocks that are surrounded by a cladding glass with a lower melting point. In order to obtain the high surface quality required for X-ray optics, the blocks are stretched into small fibres, thereby reducing the surface roughness. Several of these fibres can be assembled and fused into multifibre bundles. Etching away the glass fibre cores leads to the desired micropore structure, in which the pore walls are formed by the remaining cladding glass. The Wolter-I geometry is reproduced by thermally slumping separate multifibre plates. Glass Micropore Optics are even lighter than Silicon Pore Optics, but achieve a moderate angular resolution of about 30~arcseconds. They are especially interesting for small planetary exploration missions, but also for X-ray timing missions that require large collecting areas. The first implementation of Glass Micropore Optics on a flight programme will be in the Mercury Imaging X-ray Spectrometer on the ESA/JAXA mission BepiColombo, planned to launch in 2014 \cite{fraser}.

Currently, we are exploring the feasibility of pulsar-based navigation in the light of these new technologies. On the basis of simulated pulse profiles as measured by a virtual detector \cite{schmid} which moves along actual spacecraft trajectories (e.g., the trajectory of an interplanetary mission like Rosetta, or L2 missions like Planck and Herschel), we are studying the influence of telescope and detector parameters such as angular resolution, collecting area, temporal resolution etc. on the performance of the navigation algorithm. The aim is to find a realistic design based on an optimal configuration of these parameters and mission requirements like weight, cost, complexity and energy budget.

\begin{table}[!t]
\renewcommand{\arraystretch}{1.3}
\caption{Comparison of X-ray optics technologies (from \cite{bavdaz2}). }
\label{table1}
\centering
\begin{tabular}{|l|l|l|}
\hline
     &  Angular     & Mass per effective  \\
     &  resolution  & area (at 1 keV) \\
\hline
   Chandra    & $0.5''$ & $18\,500$ kg/m$^2$ \\
   XMM-Newton & $14''$  & $2300$ kg/m$^2$ \\
   Silicon Pore Optics & $5''$   & $200$ kg/m$^2$ \\
   Glass Micropore Optics  & $30''$  & $25$ kg/m$^2$ \\
\hline
\end{tabular}
\end{table}

\section{Summary and Outlook}
Pulsar-based navigation is an alternative to Earth-bound navigation---especially, but not exclusively, for interplanetary and deep space missions. As opposed to conventional methods of position determination by radar tracking from Earth, the accuracy of pulsar-based navigation does not depend on the distance between spacecraft and Earth, which is one of its outstanding advantages; furthermore, it supports autonomous space flight. As far as data analysis is concerned, position accuracies of $\approx$~5~kilometres can be achieved, using the optimal combination of pulsars, thereby surpassing the precision of conventional navigation for spacecraft at a distance greater than $\approx$~1~astronomical unit from Earth. 

With current telescope technology used for X-ray astronomy, pulsar-based navigation seems applicable only on large space missions. However, with the next generation of low-mass X-ray mirrors based on Silicon Pore Optics or Glass Micropore Optics, it might become technically feasible even for small-sized missions.

\section*{Acknowledgment}

MGB thanks P. Predehl and J. Tr\"umper for valuable comments and suggestions. MGB and TP acknowledge support from 
the International Max-Planck Research School on Astrophysics at the Ludwig-Maximilians University, Munich.



%

\IEEEtriggeratref{7}

\end{document}